\newif\ifJSC
\JSCfalse
\newif\ifSYNASC
\SYNASCtrue
\documentclass[conference]{IEEEtran}
\IEEEoverridecommandlockouts
\usepackage{cite}
\usepackage{amsmath,amssymb,amsfonts}
\usepackage{algorithmic}
\usepackage{graphicx}
\usepackage{textcomp}
\usepackage{xcolor}
\usepackage{url}
\def\BibTeX{{\rm B\kern-.05em{\sc i\kern-.025em b}\kern-.08em
    T\kern-.1667em\lower.7ex\hbox{E}\kern-.125emX}}
\ifJSC
\usepackage[hyphens]{url}
\fi
\usepackage{todonotes}
\usepackage{algorithm}
\usepackage{algorithmic}
\usepackage{amsmath}
\usepackage{amsthm}
\usepackage{comment}
\usepackage{xspace}
\usepackage{mathtools}
\usepackage{subcaption}
\newtheorem{definition}{Definition}

\newtheorem{remark}{Remark}

\ifJSC
\journal{J.~Symbolic Computation}
\fi

\def\RR{\mathbb{R}}

\def\res{\mathop{\rm res}\nolimits}

\newcommand{\true}{\text{\emph{true}}} 
\newcommand{\false}{\text{\emph{false}}} 

\newcommand{\VS}[3]{#1[#2\,/\hskip-2pt/\,#3]}
\newcommand{\algsubs}[3]{#1[#2\,/\,#3]}

\newcommand{\realNumbers}{\mathbb{R}}

\newcommand{\prefix}[2]{Q_{#2 - #1 + 1} x_{#2 - #1 + 1} \dots Q_{#2} x_{#2}}

\newcommand{\Qxiblock}[2]{Q_{#1} x_{#1,1},\ldots,x_{#1,#2_{#1}}}

\newcommand{\rel}[3][\rho]{#2 \, #1 ~ #3}

\newcommand{\Groebner}{Gr{\"o}bner}

\newcommand{\var}[1]{\mathrm{#1}}
\newcommand{\objmem}{\mapsto}
\newcommand{\func}[1]{#1} 
\newcommand{\objname}[1]{\texttt{#1}} 
\newcommand{\pckge}[1]{\texttt{#1}} 

\newtheorem{theorem}{Theorem}

\usepackage{color}

\begin{document}

\title{A Poly-algorithmic Approach to Quantifier Elimination\\
\thanks{
The second author's thesis was supported by EPSRC grant EP/N509589/1 and Maplesoft. The first and last authors are partially supported by EPSRC grant EP/T015713/1. The last author also thanks the partial support of Austrian Science Fund (FWF) grant P34501-N.}}

\author{\IEEEauthorblockN{1\textsuperscript{st} James H. Davenport}
\IEEEauthorblockA{\textit{Department of Computer Science} \\
\textit{University of Bath}\\
Bath, U.K. \\
Email: {\tt J.H.Davenport@bath.ac.uk}\\
ORCID 0000-0002-3982-7545}
\and
\IEEEauthorblockN{2\textsuperscript{nd} Zak P. Tonks}
\IEEEauthorblockA{\textit{Department of Computer Science} \\
\textit{University of Bath}\\
Bath, U.K.  \\
Email: {\tt Zak.P.Tonks@bath.edu}\\
ORCID 0000-0001-6019-1775}
\and
\IEEEauthorblockN{3\textsuperscript{rd} Ali K. Uncu}
\IEEEauthorblockA{\textit{RICAM, Austrian Academy of Sciences \&}\\
\textit{ University of Bath}\\
Austria \& U.K. \\
Email: {\tt aku21@bath.ac.uk}\\
ORCID 0000-0001-5631-6424}}


\maketitle

\begin{abstract}
Cylindrical Algebraic Decomposition (CAD) was the first practical means for doing Real Quantifier Elimination (QE), and is still a major method. 
Nevertheless, its complexity is inherently doubly exponential in the number of variables. Where applicable, virtual term substitution (VTS) is more effective, turning a QE problem in $n$ variables to one in $n-1$ variables in one application, and so on. Hence there is scope for hybrid methods: doing VTS where possible then using CAD.
\par
This paper describes such a poly-algorithmic implementation, based on the second author's Ph.D.  thesis. The version of CAD used is based on a new implementation of Lazard's 
method, with improvements to handle equational constraints, similar to the equational constraint handling in previous CAD algorithms.
\end{abstract}


%
\section{Introduction}
The second author's thesis \cite{Tonks2021a} contributed a new package \pckge{QuantifierElimination} for Maple that
amalgamates various aspects of research in Quantifier Elimination (QE)
to investigate the nuances of using them together. The focus of the project is a
comprehensive investigation into a poly-algorithm between Virtual Term Substitution \cite{Weispfenning1988} (VTS) and Cylindrical Algebraic Decomposition \cite{McCallumetal2019a} (CAD) 
and to enable extra efficiency and
features for QE. The package includes the first implementation of VTS
developed in collaboration with Maplesoft for Maple, the first implementation of
a Lazard-style CAD in Maple, and the first implementation of a Lazard-style
CAD with equational constraint (EC) optimisations in any context. This includes
recent research \cite{Nair2021b} on curtains in a Lazard-style CAD. 
Figure \ref{fig:7.8} shows that, on the benchmarks in \cite{Tonks2020d}, the poly-algorithmic approach is faster than purely using CAD, and that CAD benefitting from a single EC (\S\ref{sec:Equations}) is faster than CAD not treating ECs specially.
Multiple ECs are fortunately rare in the benchmark set, as the theory \cite{Nair2021b} is incomplete here.
\par
Let each $Q_i$ be one of $\forall,\exists$.
Real Quantifier Elimination problem is the following: given a statement $\Phi_0:=$
\begin{equation}\label{eq:QE}
\Qxiblock1k\cdots\Qxiblock{a+1}k\Phi(y_i,x_{i,j}),
\end{equation}
where $\Phi$ is a Boolean combination of $P_\alpha(y_i,x_{i,j})\rho_\alpha 0$, where $P_\alpha$ are polynomials and $\rho_\alpha\in\{=,\ne,<,\le,>,\ge\}$, produce a Boolean combination $\Psi$ of equalities and inequalities between polynomials $\overline P_\beta(y_i)$ which is equisatisfiable, i.e. $\Psi$ is true if and only $\Phi_0$ is true. If all the polynomials $P_\alpha(y_i,x_{i,j})$ in $\Phi(y_i,x_{i,j})$ have integer coefficients, we call $\Phi(y_i,x_{i,j})$ a Tarski formula.
\par
Having $\Psi$ means that QE problem is resolved. A partial solution (i.e. if some quantifiers are eliminated and we get
\begin{equation}\label{eq:QE2}
\Phi_1:=\Qxiblock1k\cdots\Qxiblock{b}l\hat\Phi(y_i,x_{i,j}),  
\end{equation}
with $b\le a+1$ and $l_b\le k_b$) is called an Intermediate Quantifier Elimination Result (IQER)  \cite[Definition 9]{Tonks2021a}. 
\par
Assuming $Q_i\ne Q_{i+1}$, 
$a$ is the number of \emph{alternations} of quantifiers. Let $n_0$ be the number of free variables $y_i$, and $n$ be the total number of variables $y_i$ and $x_{i,j}$.
\par
It is far from obvious that QE is even possible. 
Practical feasibility was first shown by Collins \cite{Collins1975}, using the method of Cylindrical Algebraic Decomposition (see \S\ref{sec:CAD}).

In this paper, we outline fundamentals behind the package \pckge{QuantifierElimination}. \S\ref{sec:VTS} and \S\ref{sec:CAD} give overviews of VTS and CAD. The poly-algorithmic approach is defined and discussed in \S\ref{sec:poly}. 
\ifSYNASC
\S\ref{sec:expRes} discusses the benchmarks and some experimental results. \S\ref{sec:conc} gives conclusions and future work.
\fi 

\section{Virtual Term Substitution}\label{sec:VTS}

Virtual Term Substitution (VTS) is a method to handle QE problems where the Tarski formulae are purely existentially quantified: see \cite{Weispfenning1988}. VTS eliminates quantifiers where the corresponding variables appear with low degree. Linear and quadratic elimination \ifSYNASC are in \fi\cite{Weispfenning1988,Weispfenning1997a}, and \cite{Kosta2016a} delineated elimination of variables appearing cubically. 

VTS is formulated for integral polynomials. Therefore,  we are implicitly assuming our polynomials to have integer coefficients, or at least rational coefficients, in this section. 
In particular, all constraints in the formula must be such that the associated polynomials {factor} to polynomials of at most degree 3, i.e. the limitations on degree really apply only to irreducibles. Due to the limitations on using VTS in terms of degree and the base ring of coefficients, some researchers view VTS as a preprocessing step before other algorithms for QE (see \cite{Strzebonski2022a}).
\pckge{QuantifierElimination} implements the case up to and including cubic VTS~\cite{Davenportetal2022a} for the first time in Maple. 

\ifSYNASC VTS\fi{} revolves around test-points generated from a quantified formula which describe the real root of a (potentially multivariate) polynomial in a particular variable. Such structural test-points may represent the exact substitution of the real root of some polynomial. Alternatively,
these structural test-points may feature $\infty$ or the infinitesimal $\varepsilon$, especially
because of the presence of strong relations where $\rho\in\{<,>,\neq\}$.
Where $\varepsilon$ is concerned, the root description of a polynomial must be provided, and the test-point implicitly means ``substitution of a value just less than ($-\varepsilon$)/more than ($+\varepsilon$) this real root''.

Another element of VTS is guards. Guards define a map from structural test-points to Tarski formulae, where a guard represents the conditions that must be satisfied for a structural test-point to be relevant. This offers credibility to usage of substitution points from multivariate polynomials that may otherwise vary in the number of real roots presented with respect to any one variable. Hence, VTS of a test-point into any formula is predicated on the result of its guard, and the guard must be conjoined with the result of virtual substitution.

For a given function $f(x)$ the algebraic substitution of $a$ in the place of $x$ is denoted by $\algsubs{f}{x}{a}:= f(a)$. In the same spirit we denote the virtual substitution of $a$ in $x$ by $\VS{f}{x}{a}$. 

\def\foo{Virtual Term Substitution \cite[Section 2.3]{Kosta2016a}}
\begin{definition}[\foo] \label{defn:virtual-substitution}  The virtual term substitution of a structural test-point $t$ for $x$ into a quantifier free relation $\rel{g(x,\mathbf{y})}{0}$ is $F(\mathbf{y})\coloneqq\VS{g}{x}{t}$, a quantifier free formula such that for any parameter values $\mathbf{a}\in\realNumbers^{n-1}$, if $\mathbf{a}$ satisfies the guard of $t$, then $\mathbf{a}$ satisfies $F$ if and only if $\realNumbers\models(\rel g0)(\mathbf{a},\algsubs{t}{\mathbf{y}}{\mathbf{a}})$. 
\end{definition}

If we have a non-atomic Tarski formula $\Phi(x,y)$, instead of a single relation $\rel{g(x,\mathbf{y})}{0}$, in Definition~\ref{defn:virtual-substitution}, VTS is done recursively~\cite{Kosta2016a}. 

In total, for a given existentially quantified Tarski formula of the form $\exists x \Phi(x,\mathbf{y})$, VTS works to identify a finite set $T = \{(\gamma_i, t_i)\}$ of pairs where $\gamma_i$ are guard conditions and $t_i$ are the sample test-points from each interval on the real axis and eliminates the quantifier on $x$ by the explicit correspondence\begin{equation}\label{eq:VTS}\exists x\  \Phi(x,\mathbf{y}) \Leftrightarrow \bigvee_{(\gamma_i, t_i)\in E} \left(\gamma_i \wedge \VS{\Phi}{x}{t_i}\right).\end{equation}

One can apply VTS to a universally quantified problem by negating the problem first: $\forall x\, \Phi(x) \equiv \lnot\exists x\, \lnot \Phi(x)$. 

Multivariate VTS revolves around successive substitutions of structural test-points in one variable at a time. This means that we easily inherit a canonical tree structure from substitutions of test-points within any one block of quantifiers. The branches of this tree are structural test-points, canonical tree levels correspond to one quantified variable each, and nodes are formulae owing to successive results of virtual substitution. These nodes correspond to Intermediate Quantifier Elimination Results (IQER).

\section{Cylindrical Algebraic Decomposition}\label{sec:CAD}
Cylindrical Algebraic Decomposition (CAD) was introduced by Collins \cite{Collins1975}. 
The key point of a sign-invariant CAD is that the sign of the polynomials is the same throughout each cell, so is that of the cell's sample point. Consider a cell $C$ in $n-1$ variables, i.e. all but $x_{a+1,k_{a+1}}$, with sample point $s$. Above it are various cells $C_1,\ldots,C_m$ with sample points $s_1,\ldots,s_m$, whose first $n-1$ coordinates are that of $s$. Depending on $Q=Q_{a+1}$:
\begin{description}
	\item[$Q=\exists$]If any of $s_1,\ldots,s_m$ satisfy $\Phi$, then clearly $\exists x_{a+1,k_{a+1}}\Phi$ is true at $s$ (and throughout $C$). Conversely, if no sample points satisfy $\Phi$, then by sign-invariance none of $C_1,\ldots,C_m$ satisfy $\Phi$ and hence $\exists x_{a+1,k_{a+1}}\Phi$ is false at $s$ (and throughout $C$);
	\item[$Q=\forall$]If all of $s_1,\ldots,s_m$ satisfy $\Phi$, then by sign-invariance in each cell, $\Phi$ is true at every point $(s,x_{a+1,k_{a+1}})$, and so $\forall x_{a+1,k_{a+1}}\Phi$ is true at $s$, and, again by sign-invariance, throughout $C$. Conversely, if $\Phi$ is not satisfied at some $s_i$, by sign-invariance it is not satisfied throughout $C_i$, and since $C_i$ projects onto $C$, $\forall x_{a+1,k_{a+1}}\Phi$ is false throughout $C$.
\end{description}
Hence we can determine the truth of $Q_{a+1}x_{a+1,k_{a+1}}\Phi$ on $C$, and indeed on all the cells in $\RR^{n-1}$. We then proceed inductively, determining truth down to $\RR^{n_0}$, where we have cells in the free variables alone, on each of which we know the truth of $\Phi_0$. If $TF(C_i)$ is the Tarski formula defining the cell $C_i$, then the quantifier free formula $\Psi$ can be taken as
$$
\bigvee_{i: \Phi_0(C_i) \equiv\hbox{true}} TF(C_i).
$$
\subsection{Variable Ordering}\label{sec:Variable}
The definition of QE, (\ref{eq:QE}), imposes a partial ordering $\prec$ on the variables for CAD: the $x_{a+1},i$ must be projected first, then the $x_a,i$, and so on, with the $y_i$ projected last. But the order within each block is unspecified. CAD, though, requires a total order $\overline\prec$ refining $\prec$. The question of choosing the best refinement has a long history, going back at least to \cite{Brown2003}.
There are various strategies, which can be classified by the amount of projection they do.
\begin{description}
\item[Full]Here we take all $n!$ (if there is complete freedom) possible orderings and project to $\RR^1$ for all of them, then choose ``the best''. This can either be sum of total degrees (sotd)
\cite{Dolzmannetal2004a}, or number of distinct real roots (ndrr) \cite{Bradfordetal2013a}. 
\item[Partial]Here we take all $n$ (if there is complete freedom) possible  choices for the first variable, do one projection to $\RR^{n-1}$, choose the best (via sotd) and repeat, hence doing $O(n^2)$ projections: ``greedy'' in \cite{Bradfordetal2013a}.
\item[None]Here one looks purely at the input. The first of these is in \cite{Brown2003}, known as ``Brown'', and there are variants in the \pckge{RegularChains}  \cite{ChenMorenoMaza2016a} package, and one for equational constraints in \cite{Tonks2021a}.
\end{description}
Given this choice, people have used machine learning to choose the strategy \cite{Huangetal2019a}.
The software allows the user to specify $\overline\prec$ directly (the system checks that $\overline\prec$  refines $\prec$) or one of several strategies for doing the refinement. Details are discussed in \cite[\S3.8]{Tonks2021a}, and experimental results in \cite[\S7.4.2]{Tonks2021a}, summarised in \S\ref{sec:conc-this}.

\subsection{CAD and Equations}\label{sec:Equations}
When using CAD for QE, we only need a CAD to be sign-invariant for the polynomials in the Boolean expression $\Phi$. McCallum \cite{McCallum1985b} replaced sign-invariant by the stronger condition of being order-invariant, which let him use a much smaller projection operator, but had the drawback that it could not be applied if a polynomial nullified. Then
\cite{McCallumetal2019a} justified Lazard's idea in \cite{Lazard1994}, using lex-least invariance and different forms of projection and lifting to avoid the nullification problem.

In fact we don't need any form of invariance {everywhere}: Collins~\cite{Collins1998} pointed out that if $\Phi$ took the form $(P_1=0)\land\widehat\Phi$ then we only need a CAD for the variety of $P_1$ which is sign-invariant for the polynomials in $\widehat\Phi$. In other words, the clause $P_1=0$ reduces the problem of decomposing the whole space to decomposing the variety. In this case $P_1$ is called an {equational constraint} (EC), In \cite{Bradfordetal2013b} this idea was studied further, and it is pointed out that one only needs a CAD which is invariant for the truth of $\Phi$ as a whole.
\par
The software in the second author's thesis \cite{Tonks2021a} was based on the \cite{McCallumetal2019a} version of Lazard's method, with some EC improvements as in \cite{Nair2021b}.
\subsection{Gr\"obner Bases}\label{sec:GB}
We might have multiple equational constraints, i.e. our problem $\Phi$ takes the form
\begin{equation}\label{eq:multiple}
(P_1=0)\land\cdots\land(P_k=0)\land\widehat\Phi.
\end{equation}
McCallum~\cite{McCallum2001}, in this case, suggested that (assuming all polynomials have non-zero degree in the variable to be eliminated, and are primitive), we consider $P_1=0$ as an equational constraint in $x_n$, then $R_{1,2}:=\res_{x_n}(P_1,P_2)$ as an equational constraint in $x_{n-1}$, $R_{1,2,3}:=\res_{x_{n-1}}(\res_{x_n}(P_1,P_2),\res_{x_n}(P_1,P_3))$ in $x_{n-2}$, etc.
\par
We are only interested in solutions on the variety of the $P_i$, and the precise form of the $P_i$ is irrelevant.
In particular, we could compute the Gr\"obner basis $\{Q_i\}$ of the $P_i$, and if we use a lexicographical order and the basis is triangular (as we'd expect ``in general position'', which of course we may not have), then we can use $Q_1$ in $x_n$, $Q_2$ in $x_{n-1}$ etc.: see \cite{Wilsonetal2012a}.
As pointed out in \cite{BuseMourrain2009}, $R_{1,2,3}$ is not the true multivariate resultant defined by $p_1=p_2=p_3=0$ (which would have degree $\mathcal{O}(d^3)$ by B\'ezout's theorem) but rather has degree $\mathcal{O}(d^4)$. Hence we would expect a substantial gain from the use of Gr\"obner bases \emph{once we have $\ge3$ equational constraints}. However, we see very little improvement in the accompanying benchmarks of package \pckge{QuantifierElimination} (see \cite[Figure 7.6]{Tonks2021a}). This is due to the rarity of such examples in the benchmarks used.

\subsection{Curtains}\label{sec:Curtains}
Most improvements to Collins' algorithm can encounter difficulties when some polynomials nullify, i.e. vanish over some subset of $\RR^k=\langle x_1,\ldots,x_k\rangle$.
The term curtain for the varieties where polynomials nullify was introduced in~\cite{Nairetal2020a}:
\def\foo{\cite[Definition 43]{Nair2021b}}
\begin{definition}[\foo]
        A variety $C\subseteq\RR^n$ is called a curtain if, whenever
        $(x,x_n)\in C$, then  $(x,y)\in C$ for all
        $y\in\RR$.
\end{definition}

$C$ is a curtain if it is a union of fibres of $\RR^n \to \RR^{n-1}$.
\def\foo{\cite[Definition 44]{Nair2021b}}
\begin{definition}[\foo]
        Suppose $f\in\RR[x_1,\ldots,x_n]$ and $S\subseteq\RR^{n-1}$. We say
	that $V_f$ (or $f$) has a \emph{curtain} at $S$ if for all $(\alpha_1,\ldots,\alpha_{n-1})\in S$,
	$y\in \RR$ we have $f(\alpha_1,\ldots,\alpha_{n-1},y)=0$. We call $S$ the \emph{foot} of the curtain.
	If the foot $S$ is a singleton, we call the curtain a \emph{point-curtain}.
\end{definition}

\def\foo{\textsc{cf} \cite[Definition 45]{Nair2021b}}
\begin{definition}[\foo]\label{def:implicit_explicit}
Suppose the polynomial $f\in\RR[x_1,\ldots,\allowbreak x_n]$ factorises as $f=gh$, where $g\in\RR[x_1,\ldots,x_{n-1}]$ and $g(\alpha_1,\ldots,\allowbreak \alpha_{n-1})=0$. Then the variety of $f$ is said to contain an {explicit} curtain whose foot is the zero set of $g$. 
	A curtain which does not contain (set-theoretically) an explicit curtain is said to be {implicit}, and a curtain which contains an explicit curtain but is not an explicit curtain (because $h$ itself has curtains) is said to be {mixed}. 
\end{definition}

\par
The McCallum construction~\cite{McCallum1985b} can fail to be complete in the presence of curtains. 
 Lazard lifting, as justified in \cite{McCallumetal2019a}, does not have this problem, but, as pointed out in \cite{Nair2021b,Nairetal2020a}, attempting to merge equational constraints into Lazard lifting gives problems when the equational constraints have curtains.

Algorithms 30--34 of \cite{Tonks2021a} implement the single equational constraint theory of \cite[chapters 5, 6.1]{Nair2021b}, producing a {sign-invariant} CAD of that part of $\RR^n$ where $f=0$ (the variety of $f$) when we have an equational constraint $f=0$ where $f$ has positive degree in $x_{a+1,k_{a+1}}$. The current implementation cannot handle multiple equational constraints {as such} --- of course one can ignore that they are equational.

\subsection{Partial CAD}

Partial CAD is a key optimisation for the CAD algorithm provided by Collins \& Hong \cite{CollinsHong1991} tailored completely to QE. In particular, as opposed to building every single cell, one can terminate building the CAD when a satisfactory cell (or set of cells) has been built that solves the QE problem. As an example, for a fully existentially quantified problem, Partial CAD enables the best case that one need only build one maximum level cell, for which the sample points associated to this cell are witnesses for which substitution into $\Phi$ makes it \true{}. 
The \pckge{QuantifierElimination} package has a partial CAD implementation that closely follow the ideas in~\cite{CollinsHong1991}.

Collins--Hong \cite{CollinsHong1991} describe partial CAD via the creation of child cells and propagation of truth values from those child cells towards the root where possible. In terms of the discussion above, child cells are unevaluated cells, while the propagation evaluates them. 

Child cells are constructed in a stack construction with respect to the next canonical variable for a cell. The children of the cell $c$ are subsets of $c$ in space, in the cylinder of $c$, and in some sense ``replace'' $c$ in terms of the canonical container of unevaluated cells in CAD due to the decomposition. Real algebraic numbers and functions appear; the real algebraic numbers form the static local cell bounds found from the univariate lifting polynomials for $c$ and the real algebraic functions are used to form the local cell descriptions for cells. Here one also needs to be cautious of any bounds donated from any lifting constraint.

The idea of the truth propogation is to search for meaningful truth values amongst child cells and upon finding one attempting to propagate this up the tree. 
When a cell holds a meaningful truth value, the subtree rooted at that cell can be removed from the CAD, i.e. the cells in it need not be evaluated or otherwise processed. 
The return value of truth propogation allows us to deduce the largest possible CAD subtree to remove, instead of removing pointlessly attempting to remove every subtree of that subtree on the way up.

Lifting failures are rare but nevertheless they should be handled and avoided. In \pckge{QuantifierElimination}, nullification occurrences are only relevant on polynomials related to equational constraints' nullifications, and are characterised as curtains. Where a cell $c$ has a curtain on a pivot polynomial at level $2\leq i\leq n$, lex-least semi-evaluation~\cite{BrownMcCallum2020a} is insufficient because of the lack of cross resultants between polynomials in the relevant sets. 
Hence, the polynomial intended to deduce the finer geometry to achieve lex-least invariance or even sign invariance cannot be found. 
These curtains impede our ability to construct a lex-least \& hence sign invariant CAD, and so (at the least) identifying, and otherwise avoiding or dealing with them is of interest, e.g. because sign invariance implies truth invariance on the top level formula for QE by Partial CAD. CAD in \pckge{QuantifierElimination} will check for a curtain on any child cells $c$ that undergoes stack construction by looking at the factors of the nullifying pivot polynomial.

Point curtains do not pose a problem. One theorem using equational constraints for Lazard style CAD is given as follows. The package \pckge{QuantifierElimination} identifies and recovers such curtain issues relying on this theorem.
\def\foo{After \cite[Theorem 18]{Nair2021b}}
\begin{theorem}[\foo]\label{thm:point-curtains} Let $f\in\realNumbers[x_1,\dots,x_n]$ and let $\alpha\in\realNumbers^{n-1}$. If $f$ is an EC and has a point-curtain at $\alpha$, then $P_L^{\{f\}}$, i.e. the Lazard projection of $\{f\}$ and $\{\res_{x_n}(f,g)\}$ for all $g$ in the problem, is sufficient to obtain a sign-invariant CAD.
 \end{theorem}

\begin{remark} Theorem \ref{thm:point-curtains} means that if we construct a lex-least CAD such that the only level $n-1$ curtains are point curtains, then we can actually achieve sign invariance. This does not imply Lazard invariance, but in general Lazard invariance is merely to imply sign invariance, e.g. giving truth invariance to achieve QE. 
\end{remark}

Package \pckge{QuantifierElimination} also includes the option to use multiple equational constraints without the guarantee of a correct answer. However, the curtain problems are rare~\cite{Tonks2021a}, see Section~\ref{sec:contributions}.

\section{Poly-Algorithmic Approach}\label{sec:poly}

In this section, we discuss one of the main features of the package \pckge{QuantifierElimination}. The initial exploration of the ideas and methodology was in \cite{Tonks2019b}. The reader may wish to watch an overview of the poly-algorithmic method featured in the publicly available video \cite{Tonks2020b} first recorded for ICMS 2020.

\subsection{Handling VTS IQERs} 

The intention of the poly-algorithmic approach is to use VTS as far as possible. Once it is noticed that VTS cannot complete quantifier elimination alone (due to receiving at least one ineligible {IQER} with quartic or higher degree polynomials that doesn't allow VTS's use), the poly-algorithm ships this IQER to CAD. Additionally one would like to re-purpose and reuse the CAD calculations as much as possible in this poly-algorithmic approach.

Because we can only distribute one type of quantifier through the disjunction or conjunction formed within one block, the poly-algorithm is only used within the last block of quantifiers. This is already enough to solve a huge class of problems. For example, in the context of QE for satisfiability modulo theories (SMT) and many examples alike, the problem is homogeneously quantified, so we are always in the last block. One could distribute the (remainder) of the innermost block through the canonical boolean operator for VTS onto the ineligible \objname{IQER}s. The usage of CAD on these blindly would create cylindrical formulae to quantify with the remainder of the blocks of quantifiers. However, CAD calculations would view the variables from those remaining blocks as free and would construct geometry around them, and as the CAD complexity is highly dependent on the number of variables this can lead to unnecessary and heavy calculations. Usage of the poly-algorithm in this context would hence ``double up'' on usage of variables from later blocks of quantifiers. To make matters worse, the quantifier free output of CAD in \pckge{QuantifierElimination} is generally an extended Tarski Formula, which is not appropriate as input to any QE function in \pckge{QuantifierElimination}, so we cannot even consider ``re-quantification'' of such with the later blocks of quantifiers with the current implementation.

Hence we require the state of VTS on termination to be \begin{equation} \label{eqn:vts-termination} Qx_{n-m+1}\dots Qx_{n-t}~ B\left(B_{j=1}^{k} I^{'}_j~B_{j=1}^s I_j\right)\end{equation} for $t$ the minimum level of any $I_1,\dots,I_s$, the $s>0$ ineligible {IQER}s, and $I^{'}_1,\dots,I^{'}_k$, $k\geq0$ leaf {IQER}s to be propagated with VTS, $B$ is the canonical boolean operator corresponding to the quantifier symbol $Q\in\{\exists,\forall\}$.

	The general way we handle IQERs \ifSYNASC{}is \fi as follows:
\begin{enumerate}
\item One selects an {IQER} $I$ amongst those ineligible ($I_1,\dots,I_s$), according to some metric. \pckge{QuantifierElimination} uses a metric based on depth of the selected {IQER} by default, which is
inversely proportional to its number of quantified variables.
Perform the QE $Qx_{n-m+1}\dots Qx_{n-I\objmem{}\var{level}}~I$ via Partial CAD. We retain data for this CAD, $C$.
\item If this yields a meaningful truth value we are done. Else, select the next {IQER} $I$ to solve via the chosen metric.
\par
We expect that the {IQER}s of this tree have similar boolean structure and polynomials 
By examining the polynomials from $I$ as a set, we can measure the proportion of polynomials in this {IQER} with the polynomials from the formula of the last {IQER} used in the CAD. The formula from the last {IQER} exists at the root cell for the CAD via repurposing. 

\par
If the proportion of common polynomials meets some threshold, we reuse the CAD $C$ incrementally to solve $I$. If not, we can discard $C$ and create a new CAD to solve $I$. Additionally, $I$ must contain no free variables new to $C$ for $C$ to accommodate $I$, else we must create a new CAD. However, new quantified variables are allowable.
\item We iterate this process further, and each ineligible {IQER} uses the equivalent CAD fomularion for the purposes of full QE output later. Once again, if a meaningful truth value is attained during these calculations, we are done.
\item Upon termination of the poly-algorithm, QE output for this block of quantifiers can be described by $B\left(B_{j=1}^{k} I^{'}_j~B_{j=1}^s I_j\right)$ where each $I_j$ uses its CAD output $C_j$ as its quantifier free equivalent, and each $I^{'}_j$ uses its simplified formulas. 
\end{enumerate}

Recall that VTS assumes that the quantified input formula is a Tarski formula, i.e. of integral polynomial constraints, the poly-algorithm is only applicable on Tarski formulae. One could canonicalize the root \objname{IQER} as ineligible in the case of irrational numbers, but usage of the poly-algorithm is trivial whenever the root {IQER} is ineligible --- the QE reduces to the same method to partial CAD.\\

The best case for usage of the poly-algorithm is usage of CAD on exactly one ineligible {IQER} to receive a meaningful truth value. In the case of fully (homogeneously) quantified formulae, \true{} and \false{} are the only candidates for quantifier free equivalents of {IQER}s. In this way, there is a view for the poly-algorithm to cater well to the case for fully quantified formulae. When QE is used beneath SMT solvers, the formulae are actually viewed as fully existentially quantified, and hence the poly-algorithm accommodates SMT well via finding \true{} (i.e. satisfiability) without using all {IQER}s.
\subsection{\label{sect:polyalg-var-strategy}Incremental CAD with Traversal of the VTS tree}
Depth-wise traversal of the VTS tree in the context of the poly-algorithm means that the depths of the selected {IQER}s to repurpose the held CAD are decreasing (perhaps not monotonically). An {IQER} of strictly lesser depth than the last holds strictly more variables that are quantified. It is the default choice for \pckge{QuantifierElimination} with its alternate being the breadth-wise traversal. Breadth-wise traversal fixes the ordering of all possible variables to be included in the CAD as soon as possible, at the cost of a CAD as ``large'' as possible. Depth-wise traversal aims to create as small a CAD as possible with the aims of potentially finding a meaningful truth value for the overarching quantifier $Q$, with some scope but not complete freedom for ordering future variables.

We project with respect to the last variable in the ordering for a CAD first,
and the direction of lifting is opposite to that of projection, hence the children of the root cell are with respect to the first variable in the ordering for the CAD.
In fact, it is convenient that VTS acts in the opposite direction to CAD's projection, meaning that the direction of construction of the trees of each algorithm actually coincide.
If an {IQER} to be used in CAD is of level $j$ ($0<j<m$) in a fully homogeneously quantified problem $Qx_{n-m+1}\dots Qx_n~\Phi(x_1,\dots,x_n)$, it is in the variables $x_1,\dots,x_{n-j}$. To accommodate an {IQER} of level $k > j$, we must now extend the CAD to include the variables $x_{n-j+1},\dots,x_k$. Projection on polynomials including these variables is canonical, and the produced polynomials in $x_1,\dots,x_j$ can be fed through the rest of projection incrementally by caching. 
The new variables being such that they can appear at the end of the ordering means that the lifted CAD tree need only be {extended}, and not ``{relifted}''. The existing geometry is all valid, with the admission that the cells need be repurposed in terms of the incoming Tarski formula of the {IQER} to evaluate.

One should note that, in contrast to usage of the poly-algorithm, the alternative is to pass the state of VTS on termination (\ref{eqn:vts-termination}) as a formula to CAD in a ``whole/standard'' way (as in \cite{Strzebonski2022a}). Doing so discards usage of the poly-algorithm.  One possible advantage of this non incremental approach is that CAD has all information available at the first point --- there are no restrictions in terms of variable ordering in initial CAD calls that may not cater well to later incremental calls. Furthermore, variable strategy has no requirement to have any relation to that from VTS at all.

The comparison of these different techniques applied to QE problems can be found (to be quite similar) in survival plot Figure~\ref{fig:7.8} presented in Section~\ref{sec:expRes}.
%
%
\subsection{\label{sect:poly-share}The Poly-share Criteria}
The {poly-share criteria} refers to the criteria used to decide whether to reuse a CAD to solve an incoming {IQER}.
The most reliable way to examine whether the geometry for the CAD accommodates the incoming {IQER} well is to examine the projection bases, however this has an intrinsically exponential number of polynomials spread throughout all levels of the structure. Due to the assumption that the polynomials are similar between {IQER}s, we instead suggest to take GCDs between the polynomials of the incoming {IQER} and those from the {last {IQER} used}. As this is the point where we first examine the polynomials of the {IQER}, we also ensure that no free variables are contained within the {IQER} to preclude us from reusing the existing CAD. 

\section{Experimental Results}\label{sec:expRes}
\subsection{A Comparison of the Poly-Algorithmic Approach}

We would like to give a comparison of the poly-algorithmic outcomes on the Piano Movers Problem presented in \cite{YangZeng2006}. The problem is existentially quantified:

\ifSYNASC
	\begin{equation*}
		\begin{array}{l}
\exists a\,\exists b\,\exists c\,\exists d~a^2+b^2=r^2 \,\land\, 0\leq a \,\land\, b<0 \,\land\, 1\leq c\,\land \\ d<-1\, \land c-(1+b)(c-a) =0 \,\land\, d-(1-a)(d-b) =0.
	\end{array}
	\end{equation*}
	\else
\fi

VTS can eliminate 3 out of 4 quantifiers, but the common polynomial $-{a}^{6}+{a}^{4}{r}^{2}+2\,{a}^{5}-2\,{a}^{3}{r}^{2}-2\,{a}^{4}+{a}^{2
}{r}^{2}$ appears in all resulting IQERs of level 3 (the number of quantifiers eliminated), which factors into the irreducibles $a^2 ( {a}^{4}-{a}^{2}{r}^{2}-2\,{a}^{3}+2\,a{r}^{2}+2\,{a}^{2}-{r}^{2} )$, of which one is degree 4, hence intraversible for VTS.

Here both the  case for usage of depth-wise and breadth-wise traversal of the VTS trees are the same, considering all {IQER}s formed by VTS are level 1 (the number of quantifiers to be eliminated), and the lack of a \true{}/\false{} quantifier free equivalent means every such {IQER} must be traversed.

The first IQER passed to CAD is in itself an existentially quantified QE problem (in $a$) with 6 polynomials and one free variable $r$. It results in a CAD using 10 polynomials in projection 
and 135 leaf cells on termination. 
Next {IQER} requires 4 new projection polynomials to extend the CAD, but then no new polynomials are required and only a repurposing of different boolean structures are done. 
In the first repurposing, a polynomial gets identified as an EC
, a purely semantic identification, and used to reduce the IQERs. 

In contrast, collapsing the whole VTS tree to one QE problem for CAD to traverse results in an existentially quantified disjunction. 
Equations are nested in conjunctions amongst other relations that are not equations, so CAD can deduce no equational constraints. The resulting CAD has 13 total polynomials in projection, 7 of which are for $r$, and 6 for $a$. There are 283 leaf cells on termination, with 314 created in total including parent cells.


We give a comparison of the CAD statistics used in the poly-algorithmic quantifier elimination in Table~\ref{table:piano-movers-qe-stats}.

\begin{table}
\ifSYNASC
\caption{Statistics on Piano Mover Problem.}
\fi
\label{table:piano-movers-qe-stats}
\centering
\begin{tabular}{|c|c|c|c|c|}
\hline
Method & \# Proj. Poly. & \# ECs & \# Cells & \# Leaf Cells \\
\hline
Depth/Breadh & 10 & 1 & 154 & 143 \\
Whole & 13 & 0 & 314 & 283 \\
\hline
\end{tabular}
\end{table} 

More case studies of the poly-algorithmic approach within its IQER handling can be found in~\cite{Tonks2021a}.

\subsection{Benchmarks}
There are many experimental results in \cite{Tonks2021a}, presented in survival plot style \cite{Brainetal2017a} on quite large\footnote{452 for Figure \ref{fig:7.8}, for example.}  benchmark sets, archived in \cite{Tonks2020d}. Note that this style plots, at ``200 examples'' say, the time each implementation took on the best 200 examples \emph{for it}, which may not be the same 200 examples as a different implementation uses for its plot.

All benchmarking was undertaken on a computer running Maple 2020.1 on Ubuntu 18.04.3 with 16GB of RAM and an Intel i5-4590T CPU running at 2.00 GHz.

\subsection{The CAD implementation}\label{sec:conc-this}
\begin{figure}[h]
\includegraphics[width=.48\textwidth]{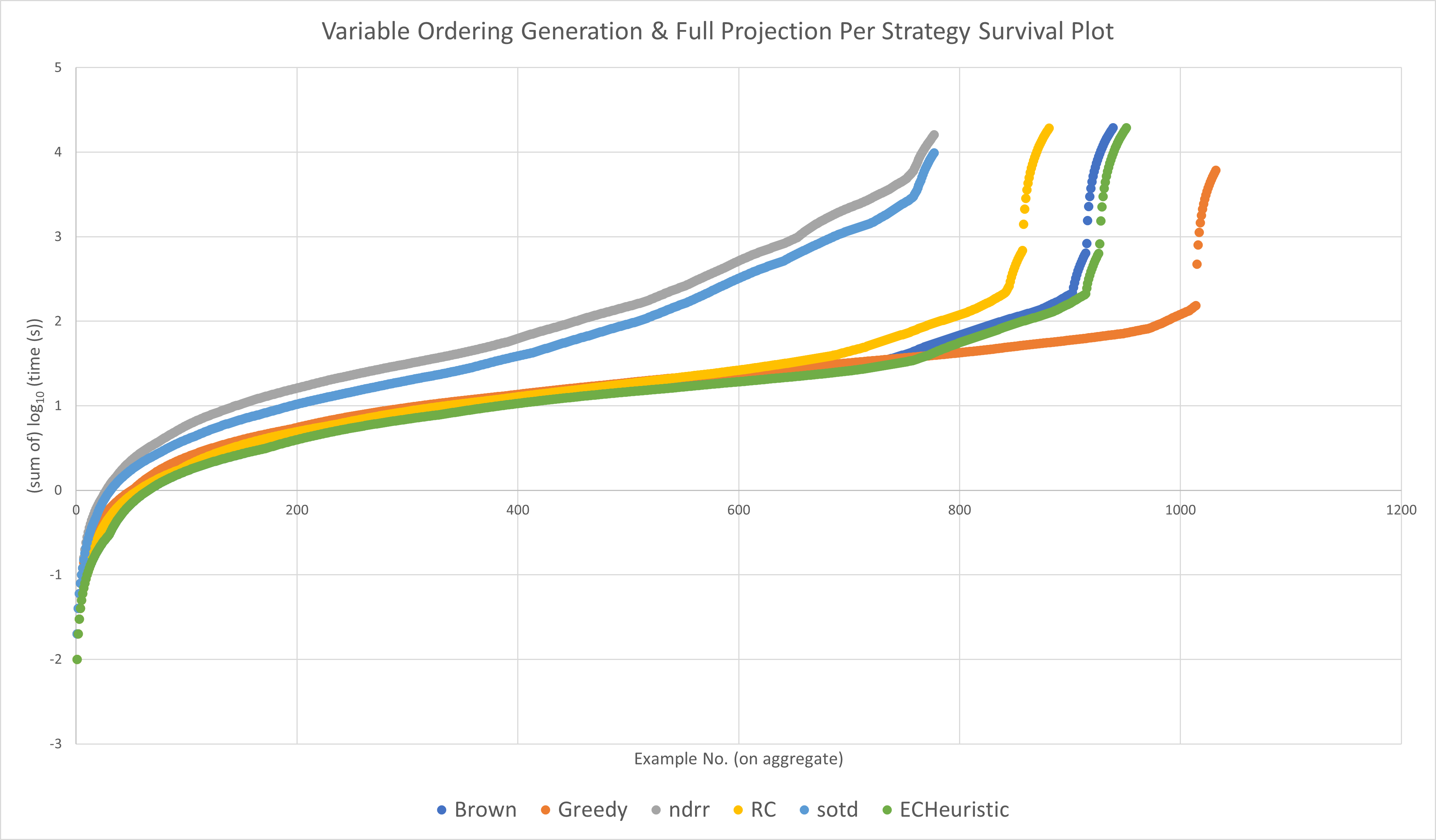}
	\caption{\cite[Figure 7.1]{Tonks2021a}\label{fig:7.1}}
\end{figure}

When a CAD with no constraints on the variable order is requested, there are many possible heuristics for choosing the ordering (see \S\ref{sec:Variable}). The benchmarks (see Figure \ref{fig:7.1}) show that the use of the greedy strategy appears to perform best here. Usage of greedy is less expensive than that of ``full sotd'' while still generating $\mathcal{O}(n)$ projection sets for each choice of $x_1$ in this case.
Considering greedy's attention to equational constraints, and \pckge{QuantifierElimination}'s sensitivity to such in terms of the fact the implementation has intended to also pay much attention to ECs, this may be unsurprising. However, we note that \pckge{QuantifierElimination} package finds usage of \Groebner{} bases preprocessing (which has already chosen a variable ordering) to be incompatible with usage of the greedy CAD variable strategy, so the timings never include generation of a \Groebner{} basis where equational constraints are concerned, which may flatter the timings for {projection} for greedy.

ECHeuristic is essentially equivalent to usage of the Brown heuristic for most examples. Overall, we see a slight improvement from ECHeuristic via the tail end of the curve, most likely as a result of the extra attention to equational constraints. The shape of the curve for \pckge{RegularChains}'~\cite{ChenMorenoMaza2016a} 
heuristic is similar to that of Brown or ECHeuristic. To conclude inspection of the strategies, one sees that ndrr and sotd live up to their $\mathcal{O}(n!)$ expectations.
\subsection{CAD benchmarking}
\pckge{QuantifierElimination} includes standalone implementations of CAD, both full CAD and partial CAD.  Although \func{CylindricalAlgebraicDecompose}, the full CAD function of the package, accepts (quantified) Real formulae, because of the intention to produce every single maximum leaf cell, it never evaluates truth values on the leaf cells it yields. It essentially decomposes the formula as the canonical sets of inequalities and equational constraints and performs full CAD. It can also accept pure sets of polynomials without relational operators, in which case evaluation of truth values is ill defined. Partial CAD call, \func{PartialCylindricalAlgebraicDecompose}, accepts an unquantified formula, and in this case produces a {CAD-Data} object, which can be examined to enumerate statistics on cells, such as the number of cells with truth value \true{}. 

The CAD benchmarking examples are rather inhomogeneous in type, some being lists of polynomials, or relations, else just unquantified formulae. The CAD computations, see \cite[Figure 7.4]{Tonks2021a}, shows that this Lazard-style implementation is distinctly better than the previous Projection--Lifting implementation (basically McCallum), but is beaten by the Regular Chains implementation. 
\cite[Figure 7.5]{Tonks2021a} shows that Regular Chains tends to produce (typically $50$-times) fewer true cells than Lazard-style, which probably accounts for much of the difference. The use of Gr\"obner bases reduces the number of cells but the reduction at hand may not be significant in many cases, see \cite[Figure 7.6]{Tonks2021a}.

\subsection{QE benchmarking}

An obvious competitor to \pckge{QuantifierEliminate} is \pckge{SyNRAC}~\cite{YanamiAnai2007}, also implemented in Maple, with both VTS and CAD. When QE in \pckge{SyNRAC} by VTS yields a formula of entirely excessive degree, \pckge{SyNRAC} switches to CAD in a non poly-algorithmic sense.

\func{PartialCylindricalAlgebraicDecompose}'s performance is close under differing usage of ECs ({single} vs. {multiple}). They act identically for examples without ECs, so some broad similarity of the curves is to be expected, but in fact these two approaches completed exactly the same number of examples. One consideration is that in the case for multiple ECs use, when multiple ECs are genuinely identified, CAD lifting must check for curtains more frequently, and in particular on larger (lower level) projection polynomials, which attributes cost. In addition Partial CAD must generally attempt avoidance of more low level curtains as a result. 

Meanwhile, the poly-algorithm via the function \func{QuantifierEliminate} outperforms the pure Partial CAD approaches. A sizeable portion of the example set tested on are the economics QE problems via the economics database. These examples are largely linear in every variable, and usually have a high number of variables. These are evidently cases where VTS is understood to outperform CAD, with CAD suffering from the number of variables especially. 
Use of \func{PartialCylindricalAlgebraicDecompose} is surprisingly competitive on linear examples considering a pure CAD approach compared to \pckge{RegularChains}, and more broadly the \pckge{QuantifierElimination} functions seem reasonably competitive on examples that are largely linear or quadratic, whether it be due to usage of VTS or surprising cases where CAD solves such examples despite numerous variables (but potentially many ECs). Recently, using Maple 2022, it was observed in \cite{Uncuetal2023a} 
that, in an SMT setting, \pckge{QuantifierElimination} was faster than \pckge{RegularChains} 14\% of the examples considered. We predict this to be due to the use of VTS and better root isolation functions implemented by the second-named author into Maple native in the meantime.

\begin{figure}[h]
\includegraphics[width=.48\textwidth]{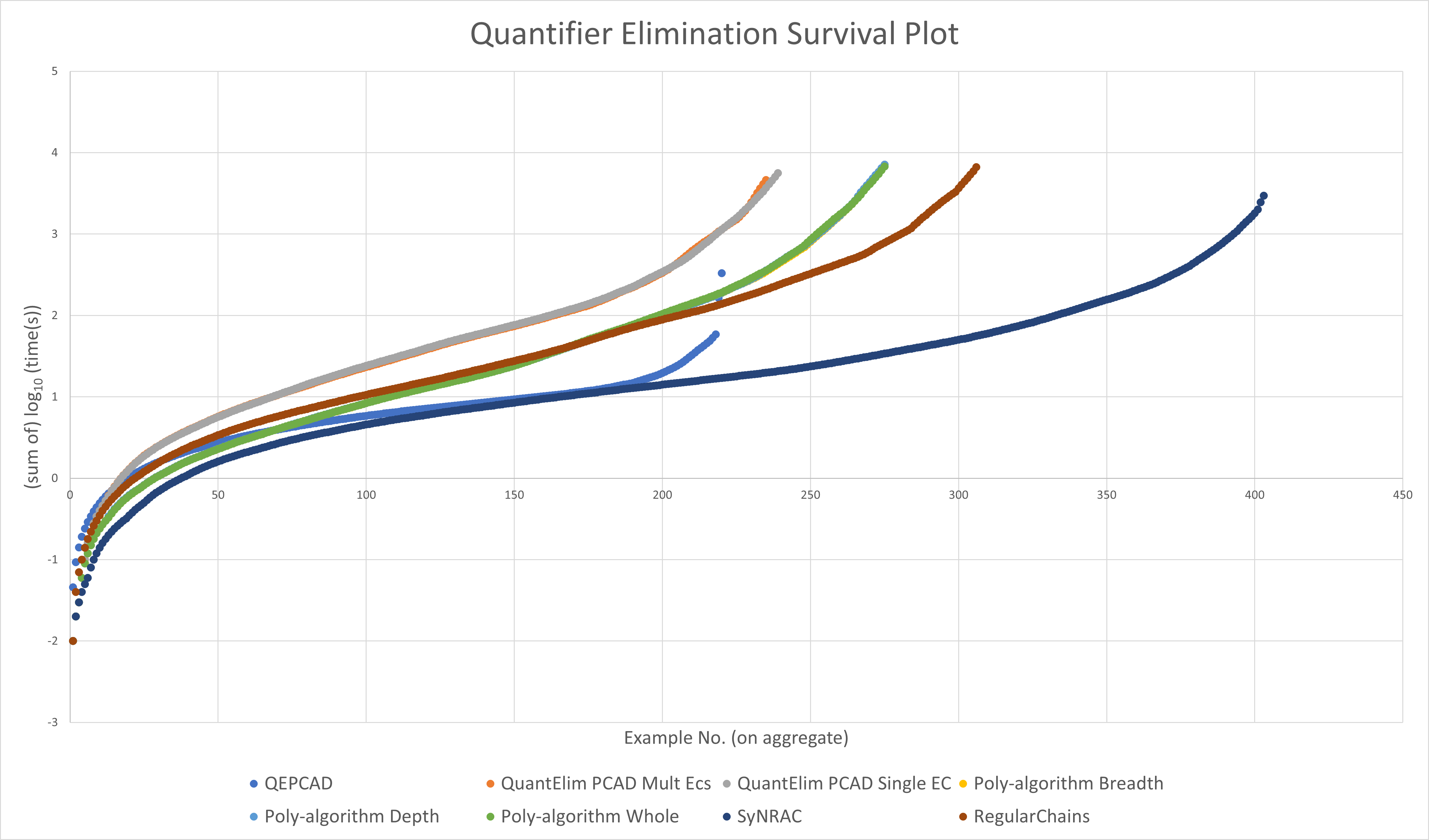}
\caption{\cite[Figure 7.8]{Tonks2021a}\label{fig:7.8}}
\end{figure}

Figure \ref{fig:7.8} evidently identifies \pckge{SyNRAC}~\cite{YanamiAnai2007} as a very competitive implementation. The Maple native \pckge{RegularChains}~\cite{ChenMorenoMaza2016a} falls not too far short, with QE being a relatively recent addition to the package, based on their CAD methodology. \pckge{QEPCAD B}~ \cite{Brown2003}  is the only software tested outside of Maple, and performs admirably. One notes timings may not be directly comparable with those generated specifically by Maple, but the usage of timeouts are always equal, and so \pckge{QEPCAD B}'s ``survival rate'' is directly comparable. Of course the low level operations used by \pckge{QEPCAD B} are completely disjoint from any of those used by the Maple packages. The shape of the curve for \pckge{QEPCAD B} largely keeps pace with that of \pckge{SyNRAC}, with a sudden sharp rise toward the later end of the curve, implying impressive performance on smaller examples, but some difficulty on the much larger ones. 

\section{Conclusions}\label{sec:conc}
\subsection{Contributions}\label{sec:contributions} 
\begin{enumerate}
\item VTS is largely a win on linear problems over CAD, including those from economics, but the Maple standard QE by CAD (\pckge{RegularChains}) can be surprisingly effective on such problems, likely due to ECs.
\item
The CAD implementation has a new feature, ``lifting constraints'', 
for problems featuring constraints that imply a
hyperrectangle in space. This can occur in biochemical applications \cite{Bradfordetal2019a}, where concentrations need to be positive.
\item Curtains are rare: the 452 examples 
only generated two curtain problems \cite[p. 280]{Tonks2021a}.
\ifSYNASC
\item Witness points can be produced by the poly-algorithm: see \cite[\S2.5]{Tonks2021a}.
\else\fi   
\end{enumerate}
\subsection{Future Work}
\begin{enumerate}
\item The work on multiple equational constraints, from \cite{Nair2021b} 
needs development. 
This theory has been translated in \cite{Davenportetal2023a} to the setting  of \cite{BrownMcCallum2020a} but not yet implemented.
\item The interaction between Gr\"obner bases (\S\ref{sec:GB}) and variable ordering (\S\ref{sec:Variable}) needs exploring. 
\item The question of when to use Gr\"obner bases with CAD is considered in \cite{Huangetal2016a}, but that was in the context of Regular Chains CAD, not a Projection--Lifting version. Hence this should be revisited.
\end{enumerate}

\bibliographystyle{ACM-Reference-Format}
\bibliography{../../../../jhd.bib}
\end{document}